\documentclass[journal]{IEEEtran}
\makeatletter
\def\ps@headings{%
\def\@oddhead{\mbox{}\scriptsize\rightmark \hfil \thepage}%
\def\@evenhead{\scriptsize\thepage \hfil \leftmark\mbox{}}%
\def\@oddfoot{}%
\def\@evenfoot{}}
\makeatother
\usepackage{amssymb}
\usepackage{amsmath}
\usepackage{bm}
\usepackage{amsthm}
\usepackage{graphicx}
\usepackage{subfigure}
\usepackage{cite}
\usepackage{enumerate}
\usepackage{mathrsfs}
\usepackage{multirow}
\usepackage{amsfonts}
\usepackage{url}
\usepackage{clrscode}
\usepackage{amsthm}
\usepackage{CJK}
\usepackage{indentfirst}
\usepackage{subfigure}
\usepackage[ruled]{algorithm2e}
\usepackage{algorithmic}

\allowdisplaybreaks[4]
\begin{document}
\title{\huge{Sensing and Communication Tradeoff Design for AoI Minimization in a Cellular Internet of UAVs}}

\author{
\IEEEauthorblockN{
\normalsize{Shuhang Zhang}\IEEEauthorrefmark{1},
\normalsize{Hongliang Zhang}\IEEEauthorrefmark{1}\IEEEauthorrefmark{2},
\normalsize{Lingyang Song}\IEEEauthorrefmark{1},
\normalsize{Zhu Han}\IEEEauthorrefmark{2}\IEEEauthorrefmark{3},
\normalsize{and H. Vincent Poor}\IEEEauthorrefmark{4}\\}
\IEEEauthorblockA{
\IEEEauthorrefmark{1} School of Electronics Engineering and Computer Science, Peking University, Beijing, 100871 China.\\
\IEEEauthorrefmark{2} Department of Electrical and Computer Engineering, University of Houston, Houston, TX 77004 USA.\\
\IEEEauthorrefmark{3} Department of Computer Science and Engineering, Kyung Hee University, Seoul 02447 South Korea.\\
\IEEEauthorrefmark{4} Department of Electrical Engineering, Princeton University, Princeton, NJ 08544 USA.}}
\maketitle
\setlength{\abovecaptionskip}{0pt}
\setlength{\belowcaptionskip}{-10pt}
\begin{abstract}
In this paper, we consider the cellular Internet of unmanned aerial vehicles~(UAVs), where UAVs sense data for multiple tasks and transmit the data to the base station~(BS). To quantify the ``freshness" of the data at the BS, we bring in the concept of the age of information~(AoI). The AoI is determined by the time for UAV sensing and that for UAV transmission, and gives rise to a trade-off within a given period. To minimize the AoI, we formulate a joint sensing time, transmission time, UAV trajectory, and task scheduling optimization problem. To solve this problem, we first propose an iterative algorithm to optimize the sensing time, transmission time, and UAV trajectory for completing a specific task. Afterwards, we design the order in which the UAV performs data updates for multiple sensing tasks. The convergence and complexity of the proposed algorithm, together with the trade-off between UAV sensing and UAV transmission, are analyzed. Simulation results verify the effectiveness of our proposed algorithm.
\end{abstract}

\section{Introduction}
Owing to the advantages of high mobility and large service coverage, unmanned aerial vehicles~(UAVs) comprise a facility that can be effectively applied in real-time sensing applications~\cite{ZSH2019}, such as air quality index monitoring~\cite{YZBSH2018}, and precision agriculture~\cite{AMCG2017}. In these applications, UAVs sense various data from different locations, and transmit the sensory data to base stations (BSs) for further processing, which comprises a cellular Internet of UAVs~\cite{ZZDS2019}.

In most of the applications in the cellular Internet of UAVs, the sensory data changes rapidly~\cite{YZBSH2018,AMCG2017}. Therefore, UAVs have to maintain the ``freshness" of the sensory data at the BS by frequent data sensing and data transmission. To measure the performance of data freshness at the BS, we bring in a new metric, i.e., Age of Information~(AoI), as proposed in~\cite{KYG2012}. The AoI is defined as the time elapsed since the most recent data update occurred, and quantifies the freshness of the sensory data, thereby converting the obscure data freshness pursuing problem into a mathematical problem that can be solved with optimization methods.

In this paper, we study a cellular Internet of UAVs, where a UAV performs data sensing and transmits the data to the BS. To keep the freshness of the data received at the BS, the UAV needs to update the sensory data frequently in a given period to minimize the total AoI of the system. Note that the success of data update is a random event and determined by the time that the UAV performs sensing and transmission. A longer time for UAV sensing increases the successful sensing probability, while a longer time for UAV transmission provides a better quality of service for communication. Given a fixed length of time, there is a trade-off between the cost of time for UAV sensing and that for UAV transmission to achieve the minimum AoI.

To minimize the total AoI of the system within a given period, it is not trivial to address the following issues. First, as the cost of time for UAV sensing and transmission involves a trade-off. The length of time that the UAV performs sensing and transmission for each task should be designed. Moreover, given the location of different tasks and their current AoI, the task scheduling, i.e., the selection of sensing task to be updated by the UAV, needs to be designed.

Sensing and transmission optimization in the cellular Internet of UAVs has been studied previously. The authors optimized the trajectory and sensing location for a set of cooperative UAVs in a cellular Internet of UAVs in~\cite{ZZDS2019'} to minimize the completion time for multiple tasks. In~\cite{WWDGH2019}, a proactive UAV path design algorithm was proposed to minimize the task completion time for the cellular Internet of UAVs. Unlike most of the existing works that consider the sensing and transmission for one task as a unit optimization objective, we further study the time consumption trade-off between the sensing and communication in the cellular Internet of UAVs.

The main contributions of this paper are summarized as follows.
First, we propose a model of the cellular Internet of UAVs, where a UAV updates the data for the sensing tasks frequently in a given period. Second, we formulate a joint sensing and transmission optimization problem to minimize the total AoI of the system, and solve the NP-hard problem with gradient descent and dynamic programming~(DP) algorithms. Third, we prove that there exists only one optimal trade-off between the time for UAV sensing and that for UAV transmission, and verify the effectiveness of our proposed algorithm with simulations.

The rest of this paper is organized as follows. In Section~\ref{System Model}, we present the system model of the Internet of UAVs. In Section~\ref{Problem Formulation}, we formulate the AoI minimization problem. The sensing and transmission optimization for one update and multiple updates are proposed in Section~\ref{Sensing and transmission optimization} and Section~\ref{Sec-Sequence}, respectively. Simulation results are presented in Section~\ref{Simulation Results}, and finally, we conclude the paper in Section~\ref{Conclusion}.
\section{System Model}\label{System Model}
In this section, we first give a brief introduction to the cellular Internet of UAVs. Afterwards, we introduce the sensing and transmission procedures of the UAV, together with the AoI of the tasks.
\subsection{Scenario Description}\label{scenario}
\begin{figure}[t]
\centerline{\includegraphics[width=3in]{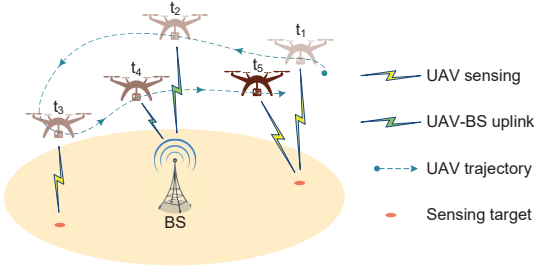}}
\caption{System model of a cellular Internet of UAVs.}
\vspace{-5mm}
\label{systemmodel}
\end{figure}

We consider a cellular Internet of UAVs as shown in Fig.~\ref{systemmodel}, which consists of one BS and one UAV. The UAV senses data of $N$ different tasks within the cell coverage, denoted by ${\cal{N}}=\{ 1, 2, ... N\}$, and transmits the sensory data to the BS for further processing. We assume that each sensing task contains one target to be sensed by the UAV. Note that the condition of a sensing target may vary frequently in time dimension. Therefore, to keep the freshness of the sensory data, the UAV needs to perform data sensing and transmission for the $N$ sensing tasks repeatedly. The UAV is required to support this sensing and transmission system for $T$ time slots.

Without loss of generality, we denote the location of the BS by $(0,0,H)$, and the location of task $n$'s sensing target by $\bm{x}_n=(x_n,y_n,0)$. In time slot $t$, let $\bm{x}(t)=(x(t),y(t),z(t))$ be the location of the UAV. Due to the space and mechanical limitations, the speed and height of the UAV satisfies
\begin{equation}\label{UAV height}\vspace{-2mm}
\|\bm{v}(t)\|\leq v_{max}, h_{min}\leq z(t)\leq h_{max}.
\end{equation}

\begin{figure}[t]\vspace{-5mm}
\centerline{\includegraphics[width=3.5in]{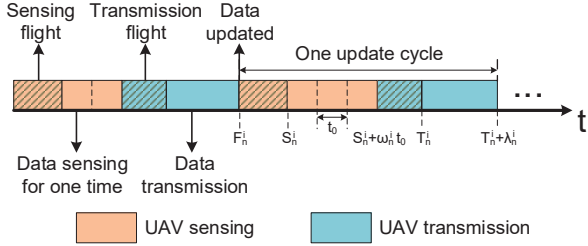}}\vspace{-3mm}
\caption{UAV sensing and transmission procedures.}
\vspace{-5mm}
\label{timeaxis}
\end{figure}

Fig.~\ref{timeaxis} illustrates the working procedure of the UAV with the time axis. We define the process that the UAV senses the data of a task and transmits the data to the BS as an update cycle. Each update cycle consists of two steps: UAV sensing and UAV transmission.
\begin{enumerate}
\item \textbf{UAV Sensing}: In UAV sensing, the UAV first moves to the location that is suitable to perform data sensing, and then senses the data of the task.
\item \textbf{UAV Transmission}: In UAV transmission, the UAV first moves to the area where the communication constraints are satisfied, and then transmits the sensory data to the BS. The UAV needs to transmit the data of a task to the BS before sensing the next one.
\end{enumerate}

\vspace{-1mm}
\subsection{UAV Sensing}\label{UAV sensing subsection}
In this subsection, we describe the UAV sensing process for data update. When sensing a task, the UAV senses data with a rate of $R_s$ in each time slot. We assume that it takes the UAV $t_0$ time slots to senses the data of a task for once. According to the probabilistic sensing model in~\cite{SK2017}, when the UAV performs data sensing for task $n$'s $i$-th update for one time, the successful sensing probability is\vspace{-2mm}
\begin{equation}\label{UAV sensing probability}\vspace{-2mm}
p_n^i(t)=e^{-\xi d_n^i(t)},
\end{equation}
where $\xi$ is a parameter evaluating the sensing performance, and $d_n^i(t)$ is the distance between the UAV and the task. Note that an unsuccessful data sensing severely delays the data update at the BS\footnote{The success of data sensing cannot be judged by the UAV immediately, and it needs to be verified by the BS with further data processing.}. Therefore, the UAV may repeatedly sense the data of a task for multiple times to achieve a higher successful sensing probability. Let $\omega_n^i$ be the number of times that the UAV senses task $n$'s data for the $i$-th update, and the successful sensing probability of this update can be shown as\vspace{-2mm}
\begin{equation}\label{sensing probability}\vspace{-2mm}
\mathscr{P}_n^i=1-(1-p_n^i(t))^{\omega_n^i}.
\end{equation}

We denote the start time of the UAV sensing for the $i$-th update of task $n$ by $F_{n}^i$. After the sensing flight, the UAV starts the data sensing in $S_n^i$, and the time to complete the data sensing can be expressed as $S_n^i+\omega_n^i t_0$. For the sake of the sensing quality, we set a minimum successful sensing probability threshold $p_{th}$ for the UAV. The successful sensing probability should satisfy
\begin{equation}\label{sensing threshold}\vspace{-2mm}
\mathscr{P}_n^i\geq p_{th}, \forall n\in\mathcal{N}.
\end{equation}

\subsection{UAV Transmission}\label{Transmission}
After UAV sensing, the UAV needs to transmit the sensory data to the BS. It is assumed that the UAV is assigned to a dedicated subchannel in the system, and thus, there is no interference in the UAV transmission process. For UAV transmission, we utilize the air-to-ground propagation model proposed in~\cite{AKJ2014}. In time slot $t$, the line-of-sight (LoS) and non-line-of-sight (NLoS) pathloss models from the UAV to the BS are given by
$PL_{L}(t)=L_{FS}(t)+20 \log(d_{UAV,BS}(t))+\eta_{LoS}$, and $PL_{N}(t)=L_{FS}(t)+20 \log(d_{UAV,BS}(t))+ \eta_{NLoS}$,
where $L_{FS}(t)$ is the free space pathloss given by $L_{FS}(t)=20\log(f)+20\log(\frac{4\pi}{c})$, $f$ is the system carrier frequency, and $d_{UAV,BS}(t)$ is the distance between the UAV and the BS. $\eta_{LoS}$ and $\eta_{NLoS}$ are additional attenuation factors due to the LoS and NLoS connections. Considering the antennas on the UAV and the BS placed vertically, the probability of LoS connection is given by
$Pr_{L}(t)=\left({1+\alpha\exp(-\beta(\phi(t)-\alpha))}\right)^{-1},$
where $\alpha$ and $\beta$ are environmental parameters, and $\phi(t)=\sin^{-1}((z(t)-H)/d_{UAV,BS}(t))$ is the elevation angle. The average pathloss in dB can then be expressed as
\begin{equation}\label{average PL}
PL_{a}(t)=Pr_{L}(t)\times PL_{L}(t)+Pr_{N}(t)\times PL_{N}(t),
\end{equation}
where $Pr_{N}(t)=1-Pr_{L}(t)$. The average received power of the BS from the UAV is given by
\begin{equation}\label{BS receive power}
P_{R}(t)={P_T}/{10^{PL_{a}(t)/10}},
\end{equation}
where $P_T$ is the transmission power of the UAV, which is considered as a fixed value in this paper\footnote{The transmission power design is independent from the proposed problem, which does not affect the optimization in this paper.}. The signal-to-noise ratio (SNR) of the transmission link in time slot $t$ is shown as $\gamma(t)=P_{R}(t)/\sigma^2$, where $\sigma^2$ is the variance of additive white Gaussian noise with zero mean. The data rate can be expressed as
\begin{equation}\label{rate}
R(t)=W_B\log_2\left(1+\gamma(t)\right),
\end{equation}
where $W_B$ is the bandwidth of the subchannel.

For the sake of transmission quality, we set an SNR threshold for the cellular link, which can be expressed as
\begin{equation}\label{transmission threshold}
\gamma(t)\geq \gamma_{th}, T_n^i\leq t\leq T_n^i+\lambda_n^i-1, \forall n\in\mathcal{N}.
\end{equation}
Let $T_n^i$ be the time slot that the UAV starts to transmit the sensory data of task $n$ for the $i$-th update, and $\lambda_n^i$ be the duration of this transmission. The received SNR constraint can be written as $\gamma(t)\geq \gamma_{th}, T_n^i\leq t\leq T_n^i+\lambda_n^i-1, \forall n\in\mathcal{N}$. To complete the data transmission for a task, the transmitted data should be no less than the sensory data, i.e.,
\begin{equation}\label{transmission constraint}
\sum_{t=T_n^i}^{T_n^i+\lambda_n^i-1}R(t)\geq R_s \omega_n^i t_0.
\end{equation}

\subsection{Age of Information}\label{AoI define Sec}
In this network, the sensory data of each task varies with time, and the validity of the data is significantly related to the timeliness of sensing and transmission. Therefore, we introduce the concept of AoI that describes the freshness of the received data at the BS, to measure the performance of the system.

We consider the $i$-th data update of task $n$ is successful at the BS side when the following two conditions are satisfied: 1)~the sensing is successfully performed by the UAV, 2)~the UAV has finished the transmission of the update data. The AoI of task $n$ is defined as the time expectation since the latest successful sensing for task $n$ is completed by the UAV. To be specific, if the data transmission for task $n$ for the $i$-th update is completed in time slot $t$, i.e., $t=T_n^i+\lambda_n^i$, it has a possibility of $\mathscr{P}_n^i$ to be updated successfully, and the corresponding AoI is denoted by $A_n(t)|_{s}=t-(S_n^i+\omega_n^i t_0)$. It also has a possibility of $1-\mathscr{P}_n^i$ to meet with an update failure, with which the AoI is given by $A_n(t)|_{f}=A_n(t-1)+1$. Therefore, the AoI of task $n$ in time slot $t$ is given as
\begin{equation}
A_n(t)\hspace{-.5mm}=\hspace{-.5mm}\left\{\hspace{-2mm}
             \begin{array}{lr}
            \mathscr{P}_n^i\times A_n(t)|_{s}\hspace{-.5mm}+\hspace{-.5mm}(1\hspace{-.5mm}-\hspace{-.5mm}\mathscr{P}_n^i)\hspace{-.5mm}\times \hspace{-.5mm} A_n(t)|_{f}, \hspace{-2mm}&\forall t\hspace{-.5mm}=\hspace{-.5mm}T_n^i\hspace{-.5mm}+\hspace{-.5mm}\lambda_n^i, \\
             A_n(t-1)+1, &\text{otherwise},
             \end{array}
\right.
\end{equation}
with $A_n(0)=0, \forall n\in \mathcal{N}$. The total AoI of task $n$ for $T$ time slots can be expressed as $A_{n}^T=\sum_{i=1}^T A_n(t)$.
\section{Problem Formulation and Decomposition}\label{Problem Formulation}
To keep the data freshness at the BS, we aim to minimize the total AoI of the $N$ tasks. Given the total time being $T$ time slots, the time for UAV sensing and UAV transmission has a trade-off. Therefore, we optimize the sensing time, transmission time, UAV trajectory, and task scheduling. The problem can be formulated as
\begin{equation} \label{system_optimization}
\begin{split}
\mathop{\min}\limits_{\substack{\textbf{$\{S_n^i\},\{\omega_n^i\}$}\\ \textbf{$\{T_n^i\},\{\lambda_n^i\}$}\\\{\bm{v}(t)\}}}& \sum_{n=1}^N A_{n}^T, \\
\textbf{\emph{s.t. }} &(\ref{UAV height}), (\ref{sensing threshold}), (\ref{transmission threshold}), \; \mbox{and}\;(\ref{transmission constraint}).
\end{split}
\end{equation}

Problem~(\ref{system_optimization}) contains both discrete variables $S_n^i,\omega_n^i, T_n^i, \lambda_n^i$ and continuous variable $\bm{v}(t)$, which is NP-hard. To solve it efficiently, we decompose problem~(\ref{system_optimization}) into two subproblems: 1) UAV sensing and transmission trade-off optimization in one update cycle, and 2) UAV sensing and transmission optimization in multiple update cycles.

\textbf{Subproblem 1: Sensing and Transmission Trade-off Optimization in One Update Cycle.} Given the total length of time, the minimization of total AoI for the $N$ tasks is equivalent to the maximization of AoI reduction. In time slot $t$, when the UAV completes UAV sensing and transmission to update the sensory data of task $n$ for the $i$-th time, the reduction of AoI can be written as $\mathscr{P}_n^i\times A_n(t)$. The expression of $A_n(t)$ can be expanded as
\begin{equation}
\begin{split}
A_n(t)=&(S_n^i+\omega_n^i t_0-S_n^{i-1}-\omega_n^{i-1} t_0)+(1-\mathscr{P}_n^{i-1})\\ \times&(S_n^{i-1}+\omega_n^{i-1} t_0-S_n^{i-2}-\omega_n^{i-2} t_0)\\+&\cdots+(1-\mathscr{P}_n^{i-1})^{i-1}\times (S_n^{1}+\omega_n^{1} t_0).
\end{split}
\end{equation}
Given that the value of $\mathscr{P}_n^{i-1}$ is close to 1, the AoI can be approximated as $A_n(t)\simeq S_n^i+\omega_n^i t_0-S_n^{i-1}-\omega_n^{i-1} t_0$. Because of the data update, the AoI of this task decrease for $\mathscr{P}_n^i\times (S_n^i+\omega_n^i t_0-S_n^{i-1}-\omega_n^{i-1} t_0)$ in all the time slots after $t$. For simplicity, we define the value of the total reduction of AoI as the \emph{AoI gain}, denoted by $G_n^i(t)=\left(\mathscr{P}_n^i\times (S_n^i+\omega_n^i t_0-S_n^{i-1}-\omega_n^{i-1} t_0)\right)\times(T-t)$.

In one update cycle, we cannot only focus on the AoI gain regardless of the length of time, since the AoI reduction and time consumption has a trade-off\footnote{A larger AoI reduction can be obtained when the UAV moves close to the sensing target or perform more times of data sensing, which cost a larger time consumption.}. Given the total length of time being $T$ time slots, the maximum total AoI reduction equals to the maximum average AoI gain in each time slot, denoted by $G_{avg,n}^i(t)=\frac{G_n^i(t)}{T_n^i+\lambda_n^i-F_{n}^i}$. In this subproblem, we study the trade-off between UAV sensing and transmission to maximize the average AoI gain in one update cycle, which is written as
\begin{equation} \label{maximize average AoI gain}
\begin{split}
\mathop{\max}\limits_{\substack{\textbf{$\{S_n^i\},\{\omega_n^i\}$}\\ \textbf{$\{T_n^i\},\{\lambda_n^i\}$}\\\{\bm{v}(t)\}}}& G_{avg,n}^i(t),\\
\textbf{\emph{s.t. }}&(\ref{UAV height}), (\ref{sensing threshold}), (\ref{transmission threshold}), \; \mbox{and}\;(\ref{transmission constraint}).
\end{split}
\end{equation}

\textbf{Subproblem 2: Sensing and Transmission Optimization in Multiple Update Cycles.} Based on the solution to subproblem 1, we aim to maximize the total AoI reduction in multiple update cycles by task scheduling in subproblem 2, and the problem can be expressed as
\begin{equation} \label{Multiple Cycle Problem}
\begin{split}
\mathop{\max}\limits_{\textbf{$\{S_n^i\}$}, \{\bm{v}(t)\}}& \sum_{t=1}^T G_{avg,n}^i(t),\\
\textbf{\emph{s.t. }}&(\ref{UAV height}).
\end{split}
\end{equation}
The sensing and transmission constraints are not considered in this subproblem, since they can be satisfied with the solution to subproblem 1. In the following, we solve the two subproblems in Sections~\ref{Sensing and transmission optimization} and~\ref{Sec-Sequence}, respectively.
\section{UAV Sensing and Transmission Trade-off Optimization in One Update Cycle}\label{Sensing and transmission optimization}
In this section, we design the sensing time, transmission time, and UAV trajectory for a given task. Since problem (\ref{maximize average AoI gain}) is still NP-hard, in the following, we decouple it into sensing optimization and transmission optimization subproblems. Note that the sensing and transmission processes are coupled, i.e., the trajectory in UAV sensing and UAV transmission processes are connected. Therefore, we propose an iterative algorithm and optimize UAV sensing and UAV transmission processes jointly.

\emph{Sensing Subproblem:} When substituting (\ref{UAV sensing probability}) and (\ref{sensing probability}) into~(\ref{maximize average AoI gain}), the UAV sensing optimization problem is written as
\begin{equation}\label{expanded average AoI gain}
\begin{split}
\mathop{\min}\limits_{T_s^f, T_s}  G_{avg,n}^i(t)=&\left[1-(1-e^{-\xi (d_n^i(t)-\bar{v}T_s^f)})^{|\frac{T_s-T_s^f}{t_0}|}\right]\times\\ &\frac{(A_n(t)+T_s+T_t)(T-t-T_s-T_t)}{T_s+T_t},\\
\textbf{\emph{s.t. }}(\ref{UAV height}) \; \mbox{and}\;(\ref{sensing threshold}).&
\end{split}
\end{equation}
where $T_s=S_n^i+\omega_n^i t_0-t$ is the length of time for UAV sensing, $T_t=T_n^i+\lambda_n^i-S_n^i-\omega_n^i t_0$ is the length of time for UAV transmission. $T_s^f=S_n^i-t$ is the length of time for sensing flight, and $\bar{v}$ is the average UAV speed during the sensing flight.

\emph{Transmission Subproblem:} When the UAV completes sensing for a task, it transmits the sensory data to the BS with minimum time consumption. The transmission optimization subproblem can be written as
\begin{equation}\label{Transmission subproblem}
\begin{split}
\mathop{\min}\limits_{\{\bm{v}(t)\}} & T_n^i+\lambda_n^i,\\
\textbf{\emph{s.t. }}&(\ref{UAV height}), (\ref{transmission threshold}), \; \mbox{and}\;(\ref{transmission constraint}).
\end{split}
\end{equation}
\subsection{Sensing Optimization}\label{Sensing Sec}
In this part, we solve the sensing optimization subproblem in (\ref{expanded average AoI gain}), and consider the UAV transmission parameters $T_n^i$ and $\lambda_n^i$ as constants. We first give a proposition for the relation between $T_s^f$ and $T_s$ in the UAV sensing step, and then design the length of $T_s$.
\subsubsection{Optimization for $T_s^f$}\label{Fly time optimization Sec}
As shown in (\ref{UAV sensing probability}), the successful sensing probability is negatively related to the distance between the UAV and sensing target of the task. Therefore, the UAV will move towards the location of the sensing target directly with the maximum speed $v_{max}$ during the flight time in UAV sensing as long as constraint (\ref{UAV height}) can be satisfied. Given the value of $t$, $T_s$, and $T_t$, the extremum value of $G_{avg,n}^i(t)$ can be obtained when $\frac{\partial G_{avg,n}^i(t)}{\partial T_s^f}=0$, Therefore, when substituting $\frac{\partial G_{avg,n}^i(t)}{\partial T_s^f}=0$ into (\ref{expanded average AoI gain}), we have
\begin{equation}\label{fianl Ts}
m \ln m\hspace{-.5mm}-\hspace{-.5mm}(1\hspace{-.5mm}-\hspace{-.5mm}m)\ln(1\hspace{-.5mm}-\hspace{-.5mm}m)\hspace{-.5mm}=\hspace{-.5mm}(1\hspace{-.5mm}-\hspace{-.5mm}m)(d_n^i(t)\xi\hspace{-.5mm}-\hspace{-.5mm}T_s\xi v_{max}),
\end{equation}
where $m=1-e^{-\xi (d_n^i(t)-v_{max}T_s^f)}, (0<m<1)$. The numerical solution of the optimal $T_s^f$ in this transcendental equation can be solved with mathematical processing software, such as MATLAB, which is denoted by $T_s^{f,opt}$.
\subsubsection{Optimization for $T_s$}\label{Sensing time Sec}
After finding the relation between $T_s^{f}$ and $T_s$, problem~(\ref{expanded average AoI gain}) becomes a function with only one variable $T_s$. For simplicity, we denote the value of $T_s^{f,opt}$ by $f(T_s)$. The objective function is written as $G_{avg,n}^i(t)=(1-(1-e^{-\xi (d_n^i(t)-v_{max}f(T_s))})^{|\frac{T_s-f(T_s)}{t_0}|})\times\frac{(A_n(t)+T_s+T_t)(T-t-T_s-T_t)}{T_s+T_t}$. By analysing the convexity of $1-(1-e^{-\xi (d_n^i(t)-v_{max}f(T_s))})^{|\frac{T_s-f(T_s)}{t_0}|}$ and $\frac{(A_n(t)+T_s+T_t)(T-t-T_s-T_t)}{T_s+T_t}$, we verify that the value of $G_{avg,n}^i(t)$ first increases with $T_s$, and then decreases with $T_s$. Therefore, there exists only one optimal solution of $T_s$ when $T_t$ is given. In other words, given the total length of time, the time for UAV sensing and UAV transmission that corresponds to the minimum AoI has only one optimal solution, which can be summarized as the following Remark.

\textbf{Remark 1:} There exists only one optimal trade-off between the time for UAV sensing and that for UAV transmission.

The optimal value can be found efficiently with the enumerating method, with the complexity of $O(T)$.
\subsection{Transmission Optimization}\label{Transmission Sec}
In this part, we solve the transmission optimization subproblem in (\ref{Transmission subproblem}), and consider the UAV sensing parameters $T_s$ and $T_s^f$ as constants.

To minimize the transmission time, the UAV moves towards the direction with the fastest uplink rate increment, i.e., the gradient of the uplink rate $\nabla R(t) = (\frac{\partial R(t)}{\partial x},$ $\frac{\partial R(t)}{\partial y}, \frac{\partial R(t)}{\partial z})$, as long as the height constraint (\ref{UAV height}) can be satisfied. The speed of the UAV is set as the maximum value, i.e., $v_{max}$ during the flight. The transmission starts when the SNR threshold~(\ref{transmission threshold}) can be satisfied, and terminates when the transmission requirement~(\ref{transmission constraint}) is satisfied.
\subsection{Algorithm Summary}
In this part, we summarize the iterative UAV sensing and transmission optimization algorithm for one update cycle, which contains iterations of sensing optimization and transmission optimization. In the sensing optimization, we first solve the relation between the UAV flight time in UAV sensing $T_s^f$ and the UAV sensing time $T_s$ as introduced in Section~\ref{Fly time optimization Sec}. Afterwards, we solve the optimal time consumption of UAV sensing $T_s$ as described in Section~\ref{Sensing time Sec}. In transmission optimization, we solve the transmission time as given in Section~\ref{Transmission Sec}. The trajectory in UAV transmission can then be obtained. Iterations of sensing optimization and transmission optimization terminates when the average AoI gain $G_{avg,n}^i$ between two consecutive iterations is below a threshold $\omega$.
\subsection{Algorithm Analysis}
In this part, we analyse the properties of the proposed algorithm. The convergency of the algorithm is proved below.

\textbf{Proposition 1:} The iterative sensing and transmission optimization algorithm is convergent.
\begin{proof}
Given the UAV transmission variables, we can obtain the optimal value of $T_s$ and $T_s^f$ as proposed in Section~\ref{Sensing Sec}. Afterwards, we utilize the gradient method to solve the optimal solution for UAV transmission as given in Section~\ref{Transmission Sec}. The average AoI gain increases with the sensing optimization and transmission optimization in each iteration. It can be known that the average AoI gain in this system has an upper bound, and cannot increase infinitely. Therefore, the iterative sensing and transmission optimization algorithm is convergent.
\end{proof}

In the following, we elaborate the impact of the sensing and transmission thresholds on the corresponding time consumptions.

\textbf{Proposition 2:} The time for UAV sensing $T_s$ increases logarithmically with the successful sensing probability threshold $p_{th}$. The time for UAV transmission $T_t$ increases lower than logarithmic with $p_{th}$.
\begin{proof}
As shown in (\ref{sensing probability}), the successful sensing probability $\mathscr{P}_n^i$ is an exponential function of the number of times for data sensing, i.e., $\omega_n^i$ increases logarithmically with $p_{th}$. Section~\ref{Fly time optimization Sec} shows that the length of flight time in UAV sensing $T_s^f$ is not affected by the value of $p_{th}$. Therefore, the time for UAV sensing $T_s=T_s^f+\omega_n^i\times t_0$ increases logarithmically with $p_{th}$. The time for UAV transmission $T_t$ is affected by the data to be transmitted $\omega_n^i R_s$, and the transmission rate $R(t)$. The value of $\omega_n^i R_s$ increases logarithmically with $p_{th}$, while the average transmission rate increases with $T_t$. Therefore, the rate of change of $T_t$ to $p_{th}$ is lower than the logarithm one.
\end{proof}

\section{UAV Sensing and Transmission Optimization in Multiple Update Cycles}\label{Sec-Sequence}
In this section, we design the order of the tasks that the UAV chooses to update in multiple update cycles to solve problem (\ref{Multiple Cycle Problem}). The problem is first converted into a knapsack problem, and then solved with the DP algorithm.

The time consumption of one update cycle is affected by its initial location, i.e., the location that the UAV completes UAV transmission for the previous task. According to the air-to-ground transmission model in~Section~\ref{Transmission}, the transmission pathloss reduces rapidly when the elevation angle is at a high level. Therefore, the locations that the UAV completes transmission for different tasks are similar. As a result, we assume that the UAV sensing and transmission in different update cycles are independent, i.e., the time consumption of each update cycle is not affected by the previous one.

The reward of UAV sensing and transmission for task $n$ for the $i$-th time can be defined as the AoI gain, which is given as $\left(\mathscr{P}_n^i\times (S_n^i+\omega_n^i t_0-S_n^{i-1}-\omega_n^{i-1} t_0)\right)\times (T-t)$. In a specific time slot, the value of $T-t$ is the same for all the tasks, and the reward of a task is its corresponding AoI gain. The cost of the UAV sensing and transmission is the time consumption. The total time $T$ is the maximum value of the time consumption. Note that this problem is more complicated than the conventional knapsack problem since the reward value is a function of the remaining time, i.e., it varies in different time slots even for the same task.

Let $\tau_i$ be the time consumption that the UAV performs sensing and transmission for task $i$. $u_i(t)$ is the action that the UAV starts to perform sensing and transmission for task $i$ in time slot $t$, and $g_i(t)$ is the corresponding reward. We denote the ordered set of all the possible actions by $\mathcal{U}=\{u_1(1), \cdots, u_1(T), \cdots u_N(1)\cdots u_N(T)\}$. We define the subset of $\mathcal{U}$ with all the elements before $u_i(t)$ as its \emph{preamble set}, denoted by $\mathcal{P}_i(t)$. Since the UAV updates the data of at most one task in each time slot, some of the elements in $\mathcal{U}$ cannot be selected simultaneously. In the following, we propose the concept of \emph{contradictory action}.

\textbf{Definition 1:} Action $u_{i_1}(t_1)$ is a \emph{contradictory action} of action $u_{i_2}(t_2)$ if $u_{i_1}(t_1)\in \mathcal{P}_{i_2}(t_2)$ and $t_1+\tau_i>t_2$. In other words, if action $u_{i_1}(t_1)$ is performed, the data update cannot be completed in time slot $t_2$, and action $u_{i_2}(t_2)$ cannot be performed.

We denote the set of contradictory actions of $u_{i}(t)$ by $\mathcal{C}_{i}(t)$, with $\mathcal{C}_{i}(t)=\{u_i'(t-\tau_{i}+1), u_i'(t-\tau_{i}+2), \cdots, u_i'(t-1)\}, \forall i' \in \mathcal{N}$. The maximum achievable reward with the action set $\mathcal{A}$ is denoted by $\mathcal{G}(\mathcal{A})$. The relation between $\mathcal{G}(\mathcal{P}_i(t))$ and $\mathcal{G}(\mathcal{P}_i(t+1))$ can be expressed as
\begin{equation}\label{recursion}
\mathcal{G}(\mathcal{P}_i(t+1))\hspace{-.5mm}=\hspace{-.5mm}\max\{\mathcal{G}(\mathcal{P}_i(t)), \mathcal{G}(\mathcal{P}_i(t)\backslash \mathcal{C}_{i}(t))+g_i(t+1)\}.
\end{equation}
The optimal solution to the task assignment problem corresponds to the maximum achievable reward of set $\mathcal{U}$, i.e., $\mathcal{G}(\mathcal{U})$. Given that $\mathcal{G}(\mathcal{P}_1(1))=g_1(1)$, the value of $\mathcal{G}(\mathcal{U})$ can be solved with the DP algorithm.

\textbf{Proposition 3:} The complexity of the task scheduling algorithm is $O(NT)$.
\begin{proof}
For each task, the number of actions is determined by the number of tasks $N$ and the total given time $T$. The number of actions in set $\mathcal{U}$ can be expressed as $O(NT)$. The value of $\mathcal{G}(\mathcal{P}_i(t+1))$ can be obtained within a constant time if the value of $\mathcal{G}(\mathcal{P}_i(t))$ and $\mathcal{G}(\mathcal{P}_i(t)\backslash \mathcal{C}_{i}(t))+g_i(t+1)$ are already solved. Each time the recursion is performed, the set of actions to be considered is subtracted to a subset of the previous one, and it will be subtracted to $\mathcal{P}_1(1))$ within $NT$ recursions. In conclusion, the complexity of the task scheduling algorithm is $O(NT)$.
\end{proof}

\vspace{-3mm}
\section{Simulation Results}\label{Simulation Results}
In this section, we evaluate the performance of the proposed algorithm. The simulation parameters are selected based on the 3GPP specifications~\cite{3GPPR12} and existing works~\cite{ZZHBS2017}. The simulation parameters are listed in Table~\ref{Simulation Parameter}.

\begin{table}[!t]
\centering
\caption{Simulation Parameters}\label{Simulation Parameter}
\begin{tabular}{|c|c|}
\hline
\textbf{Parameter} & \textbf{Value}\\
\hline
UAV maximum height $h_{max}$ & 100 m\\
\hline
UAV minimum height $h_{max}$ & 25 m\\
\hline
UAV maximum velocity $v_{max}$ & 20 m/s\\
\hline
Length of a time slot & 10 ms\\
\hline
Time for one data sensing $t_0$ & 2\\
\hline
Number of sensing tasks $N$ & 5\\
\hline
Height of the BS $H$ & 25 m\\
\hline
Sensing parameter $\xi$ & 0.01\\
\hline
Sensory data for one data sensing & 20 Mb\\
\hline
Bandwidth $W_B$ & 1 MHz\\
\hline
\end{tabular}
\end{table}

\begin{figure}[t]\vspace{-3mm}
\centerline{\includegraphics[width=3in]{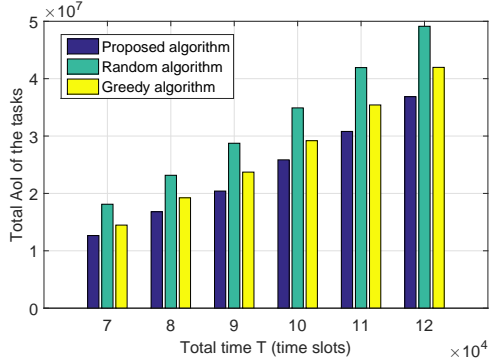}}
\caption{Total time $T$ vs. total AoI of the tasks with different algorithms.}\label{Fig1}
\vspace{-5mm}
\end{figure}
Fig.~\ref{Fig1} depicts the AoI of the system with different task scheduling algorithms. We compare the performance of our proposed algorithm with a random algorithm and a greedy algorithm. In the random algorithm, the UAV updates the data of the tasks in a random order. In the greedy algorithm, the UAV always select the sensing task with the maximum $G_{avg,n}^i$ to update its sensory data. The UAV sensing and transmission optimization method in the random algorithm and the greedy algorithm are the same as the proposed one, which is described in Section~\ref{Sensing and transmission optimization}. The AoI obtained by the proposed task scheduling algorithm is about 15\% lower than that obtained by the greedy one, and over 40\% lower than that obtained by the random one.

\begin{figure}[t]\vspace{-5mm}
    \centering
    \includegraphics[width=2.6in]{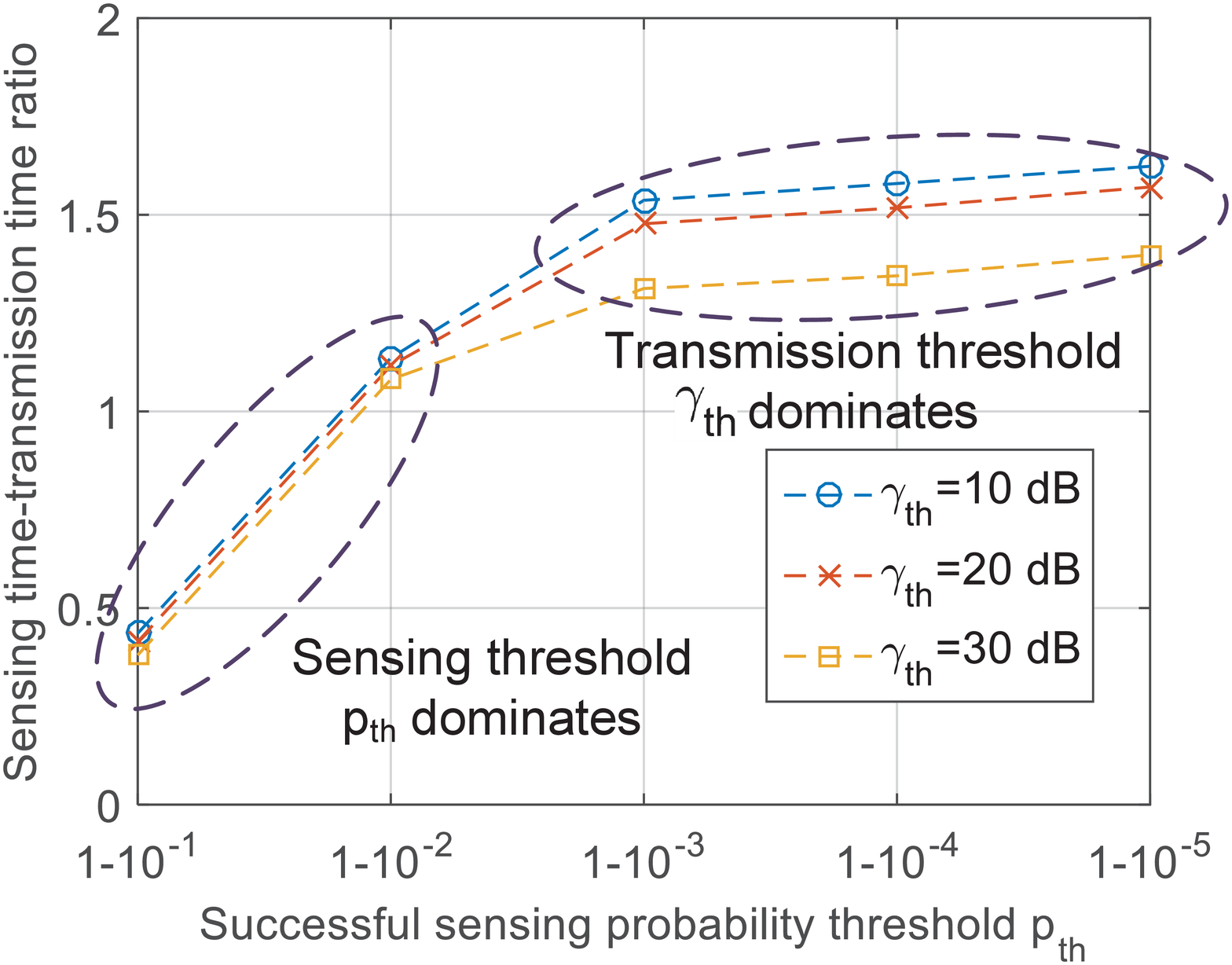}
    \vspace{-3mm}
    \caption{Successful sensing probability threshold $p_{th}$ vs. sensing time-transmission time ratio.}\label{Fig6}
    \vspace{-5mm}
\end{figure}
In Fig.~\ref{Fig6}, we plot the ratio of the UAV sensing time to the UAV transmission time with different successful sensing thresholds $p_{th}$ and transmission SNR thresholds $\gamma_{th}$. When the successful sensing threshold $p_{th}$ is less than $1-10^{-2}$, the UAV spends more time in transmission than that in sensing, and the ratio of the UAV sensing time to the UAV transmission time is mostly determined by the value of $p_{th}$. When the successful sensing threshold $p_{th}$ is larger than $1-10^{-2}$, the change of $p_{th}$ has little impact on the ratio of the UAV sensing time to the UAV transmission time. The value of the transmission SNR threshold $\gamma_{th}$ influences the ratio of the UAV sensing time to the UAV transmission time prominently.
\vspace{-5mm}
\section{Conclusion}\label{Conclusion}\vspace{-1mm}
In this paper, we have studied a cellular Internet of UAVs, where a UAV senses and transmits data from multiple tasks to the BS repeatedly for data update. We have formulated a joint sensing time, transmission time, UAV trajectory, and task scheduling optimization problem to minimize the AoI of this system within a given length of time. We have proved that there exists only one optimal trade-off between the time for UAV sensing and that for UAV transmission. Simulation results have shown that the AoI with the proposed task scheduling algorithm is about 15\% lower than that of the greedy one, and over 40\% lower than that of the random one.
\vspace{-3mm}
\section{Acknowledgement}
This paper was supported by the National Nature Science Foundation of China under grant number 61625101, grant number 61941101, and U.S. Air Force Office of Scientific Research under MURI Grant FA9550-18-1-0502.


\begin{thebibliography}{4}\vspace{-1mm}
\bibitem{ZSH2019}
H. Zhang, L. Song, and Z. Han, \emph{Unmanned Aerial Vehicle Applications over Cellular Networks for 5G and Beyond}. New York: Springer, 2019.
\bibitem{YZBSH2018}
Y. Yang, Z. Zheng, K. Bian, L. Song, and Z. Han, ``Realtime profiling of fine-grained air quality index distribution using UAV sensing," \emph{IEEE Internet Things J.,} vol. 5, no. 1, pp. 186-198, Feb. 2018.
\bibitem{AMCG2017}
B. H. Y. Alsalam, K. Morton, D. Campbell, and F. Gonzalez, ``Autonomous UAV with vision based on-board decision making for remote sensing and precision agriculture," in \emph{Proc. IEEE Aerospace Conf.,} Big Sky, MT, Mar. 2017.
\bibitem{ZZDS2019}
S. Zhang, H. Zhang, B. Di, and L. Song, ``Cellular UAV-to-X communications: design and optimization for multi-UAV networks," \emph{IEEE Trans. Wireless Commun.,} vol. 18, no. 2, pp. 1346-1359, Jan. 2019.

\bibitem{KYG2012}
S. Kaul, R. Yates, and M. Gruteser, ``Real-time status: how often should one update?" in \emph{Proc. IEEE INFOCOM,} Orlando, FL, Mar. 2012.

\bibitem{ZZDS2019'}
S. Zhang, H. Zhang, B. Di, and L. Song, ``Cellular cooperative unmanned aerial vehicle networks with sense-and-send protocol," \emph{IEEE Internet Things J.,} vol. 18, no. 2, pp. 1346-1359, Jan. 2019.
\bibitem{WWDGH2019}
H. Wang, J. Wang, G. Ding, F. Gao, and Z. Han, ``Completion time minimization with path planning for fixed-wing UAV communications," \emph{IEEE Trans. Wireless Commun.,} vol. 18, no. 7, pp. 3485-3499, Jul. 2019.

\bibitem{SK2017}
V. V. Shakhov and I. Koo, ``Experiment design for parameter estimation in probabilistic sensing models," \emph{IEEE Sensors J.,} vol. 17, no. 24, pp. 8431-8437, Dec. 2017.
\bibitem{AKJ2014}
A. Al-Hourani, S. Kandeepan, and A. Jamalipour, ``Modeling air-toground path loss for low altitude platforms in urban environments," in \emph{Proc. GLOBECOM}, Austin, TX, Dec. 2014.

\bibitem{3GPPR12}
3GPP TS 36.777, ``Enhanced LTE support for aerial vehicles," Release 15, Dec. 2017.
\bibitem{ZZHBS2017}
S. Zhang, H. Zhang, Q. He, K. Bian, and L. Song, ``Joint trajectory and power optimization for UAV relay networks," \emph{IEEE Commun. Lett.,} vol. 22, no. 1, pp. 161-164, Jan. 2018.
\end{thebibliography}
\end{document}